\documentstyle[12pt]{article}
\begin{document}
\def\ds{\displaystyle}
\def\beq{\begin{equation}}
\def\eeq{\end{equation}}
\def\bea{\begin{eqnarray}}
\def\eea{\end{eqnarray}}
\def\ve{\vert}
\def\vel{\left|}
\def\ver{\right|}
\def\nnb{\nonumber}
\def\ga{\left(}
\def\dr{\right)}
\def\aga{\left\{}
\def\adr{\right\}}
\def\lla{\left<}
\def\rra{\right>}
\def\rar{\rightarrow}
\def\nnb{\nonumber}
\def\la{\langle}
\def\ra{\rangle}
\def\ba{\begin{array}}
\def\ea{\end{array}}
\def\tr{\mbox{Tr}}
\def\ssp{{\Sigma^{*+}}}
\def\sso{{\Sigma^{*0}}}
\def\ssm{{\Sigma^{*-}}}
\def\xis0{{\Xi^{*0}}}
\def\xism{{\Xi^{*-}}}
\def\qs{\la \bar s s \ra}
\def\qu{\la \bar u u \ra}
\def\qd{\la \bar d d \ra}
\def\qq{\la \bar q q \ra}
\def\gGgG{\la g^2 G^2 \ra}
\def\q{\gamma_5 \not\!q}
\def\x{\gamma_5 \not\!x}
\def\g5{\gamma_5}
\def\sd{S_d^{cf}}
\def\su{S_u^{ad}}
\def\ss{S_s^{be}}
\def\sdp{{S}_d^{'cf}}
\def\sup{{S}_u^{'ad}}
\def\ssp{{S}_s^{'be}}
\def\sig{\sigma_{\mu \nu} \gamma_5 p^\mu q^\nu}
\def\fo{f_0(\frac{s_0}{M^2})}
\def\ffi{f_1(\frac{s_0}{M^2})}
\def\fii{f_2(\frac{s_0}{M^2})}
\def\O{{\cal O}}
\def\sl{{\Sigma^0 \Lambda}}

\title{
         {\Large
                 {\bf
The $\Sigma^0 \Lambda$ Transition Magnetic Moment 
in Light Cone QCD Sum Rules
                 }
         }
      }

\author{\vspace{1cm}\\
{\small T. M. Aliev \thanks
{e-mail: taliev@metu.edu.tr}\,\,,
A. \"{O}zpineci \thanks
{e-mail: altugoz@metu.edu.tr}\,\,,
M. Savc{\i} \thanks
{e-mail: savci@metu.edu.tr}} \\
{\small Physics Department, Middle East Technical University} \\
{\small 06531 Ankara, Turkey} }
\date{}

\begin{titlepage}
\maketitle
\thispagestyle{empty}

\begin{abstract}
Using the general form of the $\Sigma^0$ and 
$\Lambda$ currents, the $\Sigma^0 \Lambda$ transition magnetic
moment is calculated in framework of the light cone QCD sum rules.  
A comparison of our result on this quantity with the existing 
theoretical results and experimental data is presented.
\end{abstract}

~~~PACS number(s): 11.55.Hx, 13.40.Em, 14.20.Jn
\end{titlepage}

\section{Introduction}

Determination of the various fundamental parameters of hadrons from 
experiments requires information about physics at large 
distance.  Unfortunately such information can not be achieved from 
the first principles of a fundamental theory of strong interactions
because at large distance perturbation theory is invalid.  For this 
reason, to determine properties of hadrons a reliable
nonperturbative approach is needed.  Among all nonperturbative approaches, 
QCD sum rules \cite{R1} method is an  especially
powerful one in studying the properties low--lying hadrons. 
In this method, deep connection between hadron parameters and QCD vacuum
structure is established via a few condensates.  This method is adopted 
and extended in many works (see for example \cite{R2,
R3, R4} and references therein).  One of the important characteristics 
of hadrons is their magnetic moment. The nucleon magnetic moment and the
$\Sigma^0 \Lambda$ transition magnetic moment were calculated using the
external field technique in framework
of the QCD sum rules method, in
\cite{R5,R6} and \cite{R7}, respectively.

The aim of this letter is to calculate the $\Sigma^0 \Lambda$ transition 
magnetic moment in framework of an alternative approach to the
traditional sum rules method, i.e., light cone QCD sum
rules (LCQSR) (more about LCQSR method and its applications can be found 
in \cite{R8, R9} and references therein). Note that 
magnetic moment of the nucleons and decuplet baryons were studied 
in the LCQSR approach in \cite{R10} 
and \cite{R11, R12} respectively. The paper is organized as follows: 
In Sect. 2, the LCQSR for the $\Sigma^0
\Lambda$ magnetic moment is derived.  Sect. 3 is devoted to our numerical 
analysis and conclusion.

\section{Sum Rules for the $\sl$ Transition Magnetic Moment}

For the determination of the $\sl$ transition magnetic moment in LCQSR, 
we consider the following two point correlation function:

\bea
\label{1}
\Pi_\sl = i \int d^4 x e^{i p x} \la 0 \ve T\{\eta_{\Sigma^0}(x) 
\bar \eta_\Lambda (0)\} \ve 0 \ra_\gamma~,
\eea
where $T$ is the time ordering operator, $\gamma$ means external 
electromagnetic field and $\eta_{\Sigma^0}$, $\eta_\Lambda$ are
the interpolating currents with $\Sigma^0$ and $\Lambda$ quantum
numbers,respectively. 
It is well known that there is a continuum of choices for the
baryon interpolating current.  The general form of the $\Sigma^0$ and 
$\Lambda$ currents can be written as \cite{R4,R13}:
\bea
\eta_{\Sigma^0} &=& 2 \ga \eta_{\Sigma_1} + t^\prime \eta_{\Sigma_2}
\dr~, \nnb \\
\eta_{\Lambda^0} &=& 2 \ga \eta_{\Lambda_1} + t \eta_{\Lambda_2}
\dr~, 
\eea
where $t$ and $t^\prime$ are arbitrary parameters and
\bea
\eta_{\Sigma_1} &=& \frac{1}{\sqrt{2}} \epsilon_{abc} 
\left[ (u_a^T C s_b) \gamma_5 d_c + (d_a^T C s_b) \gamma_5 u_c \right]~,  \\
\eta_{\Sigma_2} &=& \frac{1}{\sqrt{2}} \epsilon_{abc} 
\left[ (u_a^T C \gamma_5 s_b) d_c + (d_a^T C \gamma_5 s_b) u_c \right]~,  \\
\eta_{\Lambda_1} &=& \frac{1}{\sqrt{6}} \epsilon_{abc} 
\left[ 2 (u_a^T C d_b) \gamma_5 s_c + (u_a^T C s_b) \gamma_5 d_c - (d_a^T 
C s_b) \gamma_5 u_c \right]~, \\
\eta_{\Lambda_2} &=& \frac{1}{\sqrt{6}} \epsilon_{abc} 
\left[ 2 (u_a^T C \gamma_5 d_b) s_c + (u_a^T C \gamma_5 s_b) d_c -
(d_a^T C \gamma_5 s_b) u_c \right] ~,
\eea
where $a$, $b$, and $c$ are color indices. Ioffe current corresponds to the
choice $t=t^\prime=-1$.

Firstly, let us discuss the hadronic representation for the correlator.  
This can be done by inserting a complete set of one hadron states
into the correlator:
\bea
\label{8}
\Pi_\sl = \sum \frac{\la 0 \ve \eta_{\Sigma^0} \ve B_1 (p_1) \ra}{p_1^2 - M_1^2}
\la B_1(p_1) \ve B_2 (p_2) \ra_\gamma
\frac{\la B_2(p_2) \ve \bar \eta_{\Lambda} \ve 0 \ra}{p_2^2 - M_1^2}~,
\eea
where $p_2 = p_1 + q$, $q$ is the photon momentum, $B_i$ form a complete set 
of baryons having the same quantum numbers as $B$ with masses $M_i$.

The interpolating current couples to the baryon states with the overlap 
amplitudes $\lambda$ defined by:
\bea
\label{9}
\la 0 \ve \eta_{\Sigma^0} \ve \Sigma^0 \ra = \lambda_{\Sigma^0} 
u_{\Sigma^0} (p)~, \nnb \\
\la 0 \ve \eta_{\Lambda} \ve \Lambda \ra = \lambda_\Lambda u_\Lambda (p)~.
\eea
It follows from Eq. (\ref{8}) that in order to write down the phenomenological 
part of the correlator, an expression for the matrix
element $\la \Sigma^0 (p_1) \ve \Lambda (p_2) \ra_\gamma$ is needed. 
This matrix element can be written as:
\bea
\label{10}
\la \Sigma^0 (p_1) \ve \Lambda (p_2) \ra_\gamma &=& \bar u (p_1) 
\left[ f_1 \gamma_\mu + i \frac{\sigma_{\mu\alpha} q^\alpha}{m_{\Sigma^0}
+ m_\Lambda} f_2 \right] u (p_2) \varepsilon^\mu~, \nnb \\
&=& \bar u (p_1) \left[ (f_1 + f_2) \gamma_\mu + \frac{(p_1 + p_2)_\mu}
{m_{\Sigma^0} + m_\Lambda} f_2 \right] u (p_2) \varepsilon^\mu~,
\eea
where the form factors $f_i$ are in general functions of $q^2 = (p_2 - p_1)^2$
and $\varepsilon^\mu$ is the polarization four vector of the photon. 
In our case, in order to evaluate the transition magnetic
moment, only the value of the form factors at $q^2=0$ are needed.  

Using Eqs. (\ref{8}), (\ref{9}), and (\ref{10}), for the phenomenological 
part of the LCQSR we get:
\bea
\label{11}
\Pi_\sl = - \lambda_{\Sigma^0} \lambda_\Lambda \varepsilon^\mu 
\frac{\not\!p_1 + m_{\Sigma^0}}{p_1^2 - m_{\Sigma^0}^2}
\left[ (f_1 + f_2) \gamma_\mu + \frac{(p_1 + p_2)_\mu}{m_\Lambda+m_{\Sigma^0}} 
f_2 \right] \frac{\not\!p_2 + m_\Lambda}{p_2^2 -
m_\Lambda^2}~.
\eea
Obviously, it follows from this expression that the correlator function contains 
a number of 
different structures.  Among all possible structures, we choose the one 
$\sim \not\!p_1\!\!\not\!\varepsilon\!\!\not\!q$ that contains the transition magnetic 
form factor $f_1 + f_2$, which when
evaluated at $q^2 = 0$ gives the transition magnetic moment in units of 
$e \hbar / (m_\Lambda + m_{\Sigma^0})$.  Isolating the
structure $\sim \not\!p_1\!\!\not\!\varepsilon\!\!\not\!p_2$ from the phenomenological 
part of the correlator, which describes the $\sl$ transition
form factor, we get
\bea
\Pi_\sl = - \lambda_{\Sigma^0} \lambda_\Lambda \frac{1}{p_1^2 - M_{\Sigma^0}^2} 
\mu_\sl \frac{1}{p_2^2 - m_\Lambda^2}~,
\eea
where $\mu_\sl = (f_1 + f_2) \ve_{q^2=0}$.

Calculation of the correlator function $\Pi_\sl$ from the QCD side leads 
to the following result:


\bea
\label{13}
\lefteqn{
\Pi_{\Sigma^0 \Lambda}(p_1^2,p_2^2) = - \frac{2}{\sqrt{3}}\,
\epsilon^{abc}\epsilon^{def}
\int d^4x e^{i p x} } \nnb \\ && 
\lla \gamma \vel  \Big\{ 
2 \g5 \sd \sup \ss \g5 + 
2 t \g5 \sd \g5 \sup \ss   \right. \right. 
\nnb \\ 
&&+ 2 t^\prime \sd \sup \g5 \ss \g5 + 
2 t t^\prime \sd \g5 \sup \g5 \ss 
\nnb \\ 
&& +
\g5 \sd \g5 \tr \su \ssp +
t \g5 \sd  \tr \su \g5 \ssp
\nnb \\ 
&&+
t^\prime \sd \g5 \tr \su \ssp \g5 +
t t^\prime \sd \tr \su \g5 \ssp \g5
\nnb \\ 
&&-
\g5 \sd \ssp \su \g5 -
t \g5 \sd \g5 \ssp \su
\nnb \\ 
&&-
t^\prime \sd \ssp \g5 \su \g5 -
t t^\prime \sd \g5 \ssp \g5 \su
\\ 
&&-
2 \g5 \su \sdp \ss \g5 -
2 t \g5 \su \g5 \sdp \ss
\nnb \\ 
&&-
2 t^\prime \su \sdp \g5 \ss \g5 -
2 t t^\prime \su \g5 \sdp \g5 \ss
\nnb \\ 
&&+
\g5 \su \ssp \sd \g5 +
t \g5 \su \g5 \ssp \sd
\nnb \\ 
&&+
t^\prime \su \ssp \g5 \sd \g5 +
t t^\prime \su \g5 \ssp \g5 \sd
\nnb \\ 
&&-
\g5 \su \g5 \tr \sdp \ss -
t \g5 \su \tr \g5 \sdp \ss
\nnb \\ 
&&-
t^\prime \su \g5 \tr \sdp \g5 \ss -
t t^\prime \su \tr \g5 \sdp \g5 \ss \left. \left. \Big\} \ver 0  \rra~,  \nnb 
\eea
where $S^{'} = C S^T C$.  Here $C$ is the charge conjugation operator and $T$ 
denotes transpose of the operator.

In order to obtain the perturbative contribution (i.e., photon is 
radiated from the freely propagating quarks) it is enough to
make the following substitution in one of the propagators in Eq. (\ref{13})
\bea
\label{14}
{S_q}^{ab}_{\alpha \beta} \rightarrow 2 \left( \int dy F^{\mu \nu} 
y_\nu {S_q}^{free} (x-y) \gamma_\mu
{S_q}^{free}(y) \right)^{ab}_{\alpha \beta}~,
\eea
where the Fock--Schwinger gauge $x_\mu A^\mu(x) = 0 $ is used and 
$S_q^{free}$ is the free quark propagator, i.e. 
\bea
S_q^{free} = \frac{i \not\!x}{2 \pi^2 x^4}~,
\eea
and the remaining two propagators are the full quark 
propagators (see below).

The expression for nonperturbative contributions can be obtained from 
Eq. (\ref{13}) by the following trick: In one of the propagators,
we made the replacement
\bea
\label{15}
{S_q}^{ab}_{\alpha \beta} \rightarrow - \frac{1}{4} \bar q^a A_j q^b (A_j)_{\alpha
\beta}~,
\eea
where $A_j = \Big\{ 1,~\gamma_5,~\gamma_\alpha,~i\gamma_5 \gamma_\alpha,
~\sigma_{\alpha \beta} /\sqrt{2}\Big\}$ and sum over $A_j$ is implied.
For the other two propagators, we substitute the full 
propagator with both perturbative and nonperturbative
contributions.

The complete light cone expansion of the light quark propagator in external 
field is calculated in \cite{R14}. It gets
contributions from the $\bar q G q$, $\bar q G G q$, $\bar q q \bar q q$ 
nonlocal operators (where $G_{\mu\nu}$ is the gluon field strength
tensor).  In the present work we consider operators with only one gluon field 
and neglect components with two gluon and four
quark fields.  Formally, neglect of the $\bar q GG q$ and $\bar q q \bar q q$ 
terms can be justified on the basis of an expansion in
conformal spin \cite{R15}.  In this approximation full light quark 
propagator is 
\bea
\label{16}
S_q &=& \frac{i \not\!x}{2 \pi^2 x^4} - \frac{\qq}{12} - \frac{m_q}{4 \pi^2 x^2} 
+ \frac{i m_q \qq}{48} \not\!x - \frac{x^2}{192} m_0^2
\qq - \frac{i m_0^2 m_q}{2^7 3^2} x^2 \not\!x \qq \nnb \\
&-& i g_s \int_0^1 dv \left[ \frac{\not\!x}{16 \pi^2 x^2} 
G^{\mu \nu}(v x) \sigma_{\mu \nu}
- v x_\mu G^{\mu \nu}(v x) \gamma_\nu \frac{i}{4 \pi^2 x^2} \right]~.
\eea
In the local part of the propagator, we neglect operators with dimension $d>5$, 
since they give negligible contribution.

It follows from Eqs. (\ref{13})--(\ref{16}) that in order to calculate QCD 
part of the sum rules we need matrix elements of nonlocal
operators between photon and vacuum state, $\la \gamma (q) \ve \bar q 
A_i q \ve 0 \ra$.  Up to twist--4, the nonzero matrix elements
given in terms of the photon wave function are \cite{R15, R16, R17}:
\bea
\la \gamma (q) \ve \bar q \gamma_\alpha \gamma_5 q \ve 0 \ra &=& \frac{f}{4} 
e_q \epsilon_{\alpha \beta \rho \sigma} \varepsilon^\beta
q^\rho x^\sigma \int_0^1 du e^{i u qx} \psi(u)~, \nnb \\
\la \gamma (q) \ve \bar q \sigma_{\alpha \beta} q \ve 0 \ra &=& 
i e_q \qq \int_0^1 du e^{i u q x} \Bigg\{ (\varepsilon_\alpha q_\beta -
\varepsilon_\beta q_\alpha) \Big[ \chi \phi(u) + x^2 \Big(g_1(u) - g_2(u)\Big) \Big]  \nnb \\
&+& \Big[ qx (\varepsilon_\alpha x_\beta - \varepsilon_\beta x_\alpha) + 
\varepsilon x (x_\alpha q_\beta - x_\beta q_\alpha) \Big] g_2
(u) \Bigg\}~,
\eea
where $\chi$ is the magnetic susceptibility of the quark condensate, $e_q$ is 
the quark charge, the functions $\phi(u)$ and $\psi(u)$
are the leading twist--2 photon wave functions, while $g_1(u)$ and $g_2(u)$ are 
the twist--4 functions.  Note that in the calculations,
the masses of the $u$ and $d$ quarks are neglected and only the terms linear 
in the strange quark mass are  taken into
account.  Therefore, under SU(2) symmetry $u$ and $d$ quark propagators and 
their condensates are 
identical, i.e., $\qu=\qd$ (the difference in the
wave functions are due to their charges only).

Substituting the photon wave functions and the expression for the quark
propagator into Eq. (\ref{13}), we can calculate the theoretical part, i.e.,
OPE part of the correlator (\ref{1}). The sum rules is obtained by equating
the phenomenological and theoretical parts of the correlator. 
In order to suppress the contributions of the higher states and the
continuum, we perform double Borel transformations on the variables
$p_1^2=p^2$ and $p_2^2=(p+q)^2$ (for more details see
\cite{R11,R12,R18,R19}), and we get the
following result for the transition magnetic moment


\bea
\label{18}
\lefteqn{
\sqrt{3}\, \lambda_\Lambda \lambda_{\Sigma^0} \,\mu_{\Sigma^0\Lambda}\, 
e^{-\ga \frac{M_\Lambda^2}{M_1^2} + \frac{M_{\Sigma^0}^2}{M_2^2} \dr} =
\ga e_d - e_u \dr \Bigg\{ - \frac{m_s}{16 \pi^2} \qq 
\Big[ t^\prime + t (-3 + 2 t^\prime) \Big]  M^4 E_1(x)  \chi \varphi(u_0)
} \nnb \\ 
&& + \frac{1}{16 \pi^2} (2 + t + t^\prime + 2 t t^\prime)   M^4 E_1(x) f
\psi(u_0) +
\nnb \\ &&
+\frac{m_s}{2 \pi^2} \qq \Big[ t^\prime + t (-3 + 2 t^\prime)\Big]  
\Big[ g_1(u_0) - g_2(u_0) \Big] M^2 E_0(x)
\nnb \\ &&
+\frac{1}{6} \qq \Big[ (-3 t + t^\prime + 2 t t^\prime) \qs + 2 (-1 + t -
t^\prime + t t^\prime) \qq
\Big]  M^2 E_0(x) \chi \varphi(u_0)
\nnb \\ &&
- \frac{2}{3} \qq \Big[ (-3 t + t^\prime + 2 t t^\prime) \qs + 2 (-1 + t -
t^\prime + t t^\prime) \qq
\Big]  \Big[ g_1(u_0) - g_2(u_0)\Big]
\nnb \\ &&
+ \frac{m_s}{12} \Big[ (2 + t + t^\prime + 2 t t^\prime) \qs + (2 + t +
t^\prime - 4 t t^\prime) \qq
\Big]  f \psi(u_0)
\nnb \\ &&
- \frac{1}{144} m_0^2 \qq   \Big[ (-4 - 11 t + 5 t^\prime + 10 t t^\prime) \qs + 8 (-1 +
t - t^\prime + t t^\prime) \qq
\Big] \chi \varphi(u_0)
\nnb \\ &&
-\frac{1}{32 \pi^4} (2 + t + t^\prime + 2 t t^\prime) M^6 E_2(x) +
\frac{1}{6} \qs \qq (2 + t + t^\prime - 4 t t^\prime)
\nnb \\ &&
- \frac{m_s}{8 \pi^2} M^2 \Big[ (2 + t + t^\prime + 2 t  t^\prime) \qs + (2 + t + t^\prime - 4 t t^\prime) \qq \Big]
\nnb \\ &&
+ \frac{m_0^2 m_s}{96 \pi^2} \Big[ 2 (2 + t + t^\prime + 2 t  t^\prime) \qs +3 (2 + t + t^\prime - 4 t t^\prime) \qq \Big]
\nnb \\ &&
- \frac{3 m_0^2}{32 \pi^2} m_s \qq (1 - t t^\prime) \left(\gamma_E - \ln \frac{M^2}{\Lambda^2} \right) \Bigg\}
\nnb \\ &&
- e_s \frac{m_0^2}{96 \pi^2} m_s \qq (1 - t t^\prime) \left(\gamma_E - \ln \frac{M^2}{\Lambda^2} \right) ~, 
\eea
where 
\bea
E_n (x) = 1 - e^{-x} \sum_0^n \frac{1}{k!} x^k~, \nnb
\eea
are the functions used to subtract the continuum, $x=s_0/M^2$, $s_0$ is the
continuum threshold and
\bea
u_0 = \frac{M_2^2}{M_1^2+M_2^2}~,~~~~~~M^2=\frac{M_1^2 M_2^2}{M_1^2+M_2^2}~,
\nnb
\eea
where $M_1^2$ and $M_2^2$ are the Borel parameters in $\Sigma^0$ and
$\Lambda$ channels. Since masses of the $\Sigma^0$ and $\Lambda$ are very
close to each other, we will set $M_1^2=M_2^2=2 M^2$, hence $u_0=1/2$. 
It follows from Eq. (\ref{18})
that in determining the $\Sigma^0\,\Lambda$ transition matrix moment one
needs to know the residues $\lambda_\Lambda$ and $\lambda_{\Sigma^0}$. These
residues are determined from baryon mass sum rules \cite{R4,R13}.  


\bea
\label{19}
M_\Lambda \lambda_\Lambda^2 e^{- \frac{M_\lambda^2}{M^2}} &=&
\frac{m_s}{192 \pi^4} (-13 + 2 t + 11 t^2) M^6 E_2(x) \nnb \\ 
&+& \frac{1}{48 \pi^2} (1-t) \left\{ (13 + 11 t) \qs + 2 (1 + 5 t) 
\qq \right\} M^4 E_1(x) \nnb \\ 
&-& \frac{m_0^2}{16 \pi^2} (1-t^2) (2 \qs + \qq) M^2 E_0(x)\nnb \\ 
&+&\frac{m_s}{18} \qq \left\{ (1 + 4 t -5 t^2) \qs + 3 (5 + 2 t + 5 t^2) \qu
\right\}~,
\eea


\bea
\label{20}
\lambda_\Lambda^2 e^{-\frac{M_\Lambda^2}{M^2}} &=& 
\frac{1}{256 \pi^4} (5 + 2t + 5t^2) M^6 E_2(x) \nnb \\ 
&+& \frac{1}{72} (1 - t) m_0^2  \qq  \left\{ 8 (1 + 2 t) \qs + 
(25 + 23 t) \qq \right\} \frac{1}{M^2} \nnb \\ 
&+& \frac{m_s}{96 \pi^2} \left\{ 3 (5 + 2 t + 5 t^2 ) \qs + 
4 (1 + 4 t - 5 t^2 ) \qq \right\} M^2 E_0 (x) \nnb \\ 
&+& \frac{m_s}{16 \pi^2} m_0^2 \qq (1 - t^2 ) 
\left\{\gamma_E - \ln \left(\frac{M^2}{\Lambda^2} \right) \right\} \nnb \\ 
&-& \frac{1}{18} \qq (1 - t) \left\{ 2 (1 + 5 t) \qs + 
(13 + 11 t) \qq \right\} \nnb \\ 
&-& \frac{m_s}{96 \pi^2} m_0^2 \left\{ (5 + 2 t + 5 t^2 ) \qs + (-5 + 4 t +
t^2 ) \qq \right\}~,
\eea


\bea
\label{21}
M_{\Sigma^0} \lambda_{\Sigma^0}^2 e^{-\frac{M_\Lambda^2}{M^2}} &=&
\frac{m_s}{64 \pi^2} (1-t^\prime)^2 M^6 E_2(x) \nnb \\ 
&+& \frac{1}{16 \pi^2} (1 - {t^\prime}) \left\{ (-1 + {t^\prime}) \qs + 
6 (1 + {t^\prime}) \qq) \right\} M^4 E_1(x) \nnb \\ 
&-& \frac{3}{16 \pi^2} m_0^2 \qq (1 - {t^\prime}^2 ) M^2 E_0(x) \nnb \\ 
&+& \frac{m_s}{6} \qq \left\{ -3 (-1 + {t^\prime}^2 ) \qs + (5 + 2 {t^\prime} + 5 {t^\prime}^2 ) \qq
\right\}~,
\eea 


\bea
\label{22}
\lambda_{\Sigma^0}^2 e^{-\frac{M_\Lambda^2}{M^2}} &=&
\frac{1}{256 \pi^4} (5 + 2 {t^\prime} + 5 {t^\prime}^2 ) M^6 E_2(x) \nnb \\ 
&+& \frac{m_s}{32 \pi^2} \left\{ (5 + 2 {t^\prime} + 5 {t^\prime}^2 ) \qs  - 
12 (-1 + {t^\prime}^2 ) \qq \right\}  M^2 E_0(x) \nnb \\ 
&+& \frac{1}{24} m_0^2 \qq (1 - {t^\prime})  \left\{ 12 (1 + {t^\prime}) \qs + 
(-1 + {t^\prime}) \qq \right\} \frac{1}{M^2} \nnb \\ 
&+& \frac{3 m_s}{16 \pi^2} m_0^2 \qq (1 - {t^\prime}^2 ) \left\{\gamma_E - 
\ln \left(\frac{M^2}{\Lambda^2} \right) \right\} \nnb \\ 
&-& \frac{m_s}{96 \pi^2} m_0^2 \left\{ (5 + 2 {t^\prime} + 5 {t^\prime}^2 ) 
\qs - 3 (-1 + {t^\prime}^2 ) \qq \right\} \nnb \\ 
&-& \frac{1}{6} \qq (1 - {t^\prime}) \left\{ 6 (1 + {t^\prime}) \qs + (-1 + {t^\prime}) \qq \right\}~,
\eea
where $\Lambda$ is the QCD scale parameter and it is chosen to be
$\Lambda=0.5~GeV$.

In this set of expressions, Eqs. (\ref{19}), (\ref{21}) and
(\ref{20}), (\ref{22}) correspond to the structures proportional to the unit
operator and $\not\!p$, respectively. In order to obtain the transition
magnetic moment $\mu_{\Sigma^0\Lambda}$, we substitute the values of
$\lambda_\Lambda$ and $\lambda_{\Sigma^0}$ from Eqs. (\ref{20}) and (\ref{22})
into Eq. (\ref{18}).

\section{Numerical analysis}
In this section we present the numerical analysis of the sum rules for the 
$\Sigma^0 \, \Lambda$ transition matrix element which has already been 
obtained in the
previous section. As can obviously be seen from Eq. (\ref{18}) the main 
input parameters of the sum rules are photon wave functions. It was shown in 
\cite{R15,R16} that the leading photon wave functions receive only small
corrections from the higher conformal spin, so that their deviation from
asymptotic form is inessential. We shall use the following expressions for
the photon wave functions \cite{R15,R17}:
\bea
\phi(u) &=& 6 u (1-u)~,~~~~~\psi(u) = 1~, \nnb \\
g_1(u) &=& - \frac{1}{8}(1-u)(3-u)~,~~~~~g_2(u) = -\frac{1}{4} (1-u)^2~.\nnb
\eea
The values of input parameters that are used in the numerical calculations
are:
$f=0.028~GeV^2$, $\chi=-4.4~GeV^{-2}$ \cite{R20} (in  \cite{R21} it is
estimated to be $\chi=-3.3~GeV^{-2}$), $\qq(1 ~ GeV)=-(0.243)^3~GeV^3$, and
$m_0^2=(0.8\pm0.2)~GeV^2$ \cite{R22}, $m_s(1 ~ GeV) = (150 \pm 50) \, MeV$, 
$\qs(1 ~ GeV)=0.8 \qq(1 ~ GeV)$.

In further numerical analysis we set $t=t^\prime$. Since the transition
magnetic moment is a physical quantity it must be independent of the
parameters $t$, continuum threshold $s_0$ and Borel mass $M^2$. So the main problem is to find a
region where the result of the transition magnetic moment is practically
independent of the parameters $t$, $s_0$ and $M^2$. 

In Fig. (1) we present the dependence of the transition
magnetic moment on the Borel parameter $M^2$. The continuum threshold $s_0$ is determined
from mass sum rules for $\Sigma$ and $\Lambda$ baryons \cite{R3, R14} and from $\Sigma^0-\Lambda$ 
mass difference sum rules \cite{star}.
It follows from this figure that the working region of Borel mass $M^2$ is 
$1~GeV^2 \le M^2 \le 1.5~GeV^2$.
Moreover we see that for the choices $s_0=3~GeV^2$ and $s_0=4~GeV^2$, the
variation in the results is about $10\%$, i.e., the transition magnetic
moment can be said to be practically insensitive to the value of the 
continuum threshold at $t=-3$, $t=-2$ and $t=3$. The result also seems to
be practically independent of the choice of the value of the parameter $t$.

Before determining the transition magnetic moment,
the next problem to be considered is to find an
appropriate region of $t$. For this purpose we have used the mass sum 
rules (see Eqs. (\ref{19})-(\ref{21})). Two criteria should be met by
the mass sum rules.  First of all, each sum rule must separately be positive.
After an analysis of the mass sum rules, we found that the
region $-0.6 \le t \le 0.9$ is unphysical for $\Lambda$, and
$-0.4 \le t \le 0.9$ is unphysical for $\Sigma^0$.
The second criteria is that the predicted mass of the baryons, obtained by
considering the ratio of the chiral odd and the chiral even mass sum
rule, should be stable in regard to variations of the parameter $t$.  In the 
early analysis of the mass sum rules, only the ratio is considered and it was 
found that the ratio stabilizes at $t=-0.2$, but as we have already noted, this value is not in the physical region.
Our analysis shows that, the most appropriate value of $t$ is given by $-0.5 \le \cos\theta \le 0.5$
where $\theta$ is defined through the relation $t=\tan\theta$.  This region of $\theta$ corresponds to 
$t \ge 1.7$ or $t \le -1.7$.  One should also note that the Ioffe current, which corresponds to the 
choice $t=-1$, is also not in the appropriate region.

In Fig. (2) the dependence of the transition magnetic moment on $\cos\theta$
at $M^2=1~GeV^2$, $s_0=3~GeV^2$ and $s_0=4~GeV^2$ is depicted. 
A reasonable value for the transition magnetic moment must
be obtained far from the unphysical region. We observe from Fig. (2) that
$\mu_{\Sigma^0 \Lambda}$ is quite stable in the region $-0.5\le \cos\theta
\le -0.5$ and practically seems to be independent of $\cos\theta$ (or $t$) and
continuum threshold $s_0$. In this region of interest we obtain
\bea
\ve \mu_{\Sigma^0 \Lambda} \ve = (1.6 \pm 0.3) \, \mu_N~, \nnb
\eea
for the transition magnetic moment,
where the errors are mainly due to variations with respect to
the Borel mass $M^2$, the continuum threshold $s_0$, the value of the parameters $\chi$, 
and omitted higher twist photon wave-function contributions.

Finally let us compare our result on $\mu_{\Sigma^0\Lambda}$  
with the results of the existing
theoretical calculations and experimental data.
For the transition 
magnetic moment $\mu_{\Sigma^0 \Lambda}$ the traditional QCD sum rules
predicts $\vel \mu_{\Sigma^0 \Lambda} \ver =1.5~\mu_N$ \cite{R7}. The  
constituent quark model predicts that $\mu_{\Sigma^0 \Lambda}=
(\mu_d-\mu_u)/\sqrt{3}$,
and with $\mu_d=-0.972~\mu_N$ and $\mu_u=1.852~\mu_N$, this result leads 
to $\mu_{\Sigma^0 \Lambda}\simeq -1.65$.
The experimental result of the transition magnetic moment is measured to be 
$\mu_{\Sigma^0 \Lambda} = - 1.6~\mu_N$ (see \cite{R23}). When we compare
these results we see that our prediction on the transition magnetic moment
is in a good agreement with those predicted by the traditional QCD sum rules
and constituent quark model, as well as with the existing experimental data.

\newpage
\section*{Figure captions}
{\bf Fig. (1)} The dependence of the transition magnetic moment
$ \ve \mu_{\Sigma^0\Lambda} \ve $ on the
Borel mass $M^2$ at $t=-3;~t=-2.0;~t=+3$ and at the continuum threshold
$s_0=3.0~GeV^2$ and $s_0=4.0~GeV^2$.\\ \\
{\bf Fig. (2)} The dependence of the transition magnetic moment
$\ve \mu_{\Sigma^0\Lambda}\ve $ on $\cos\theta$ at $M^2=1~GeV^2$, $s_0=3~GeV^2$ and
$s_0=4~GeV^2$.  

\newpage

\begin{figure}
\vskip 0.5cm
    \includegraphics{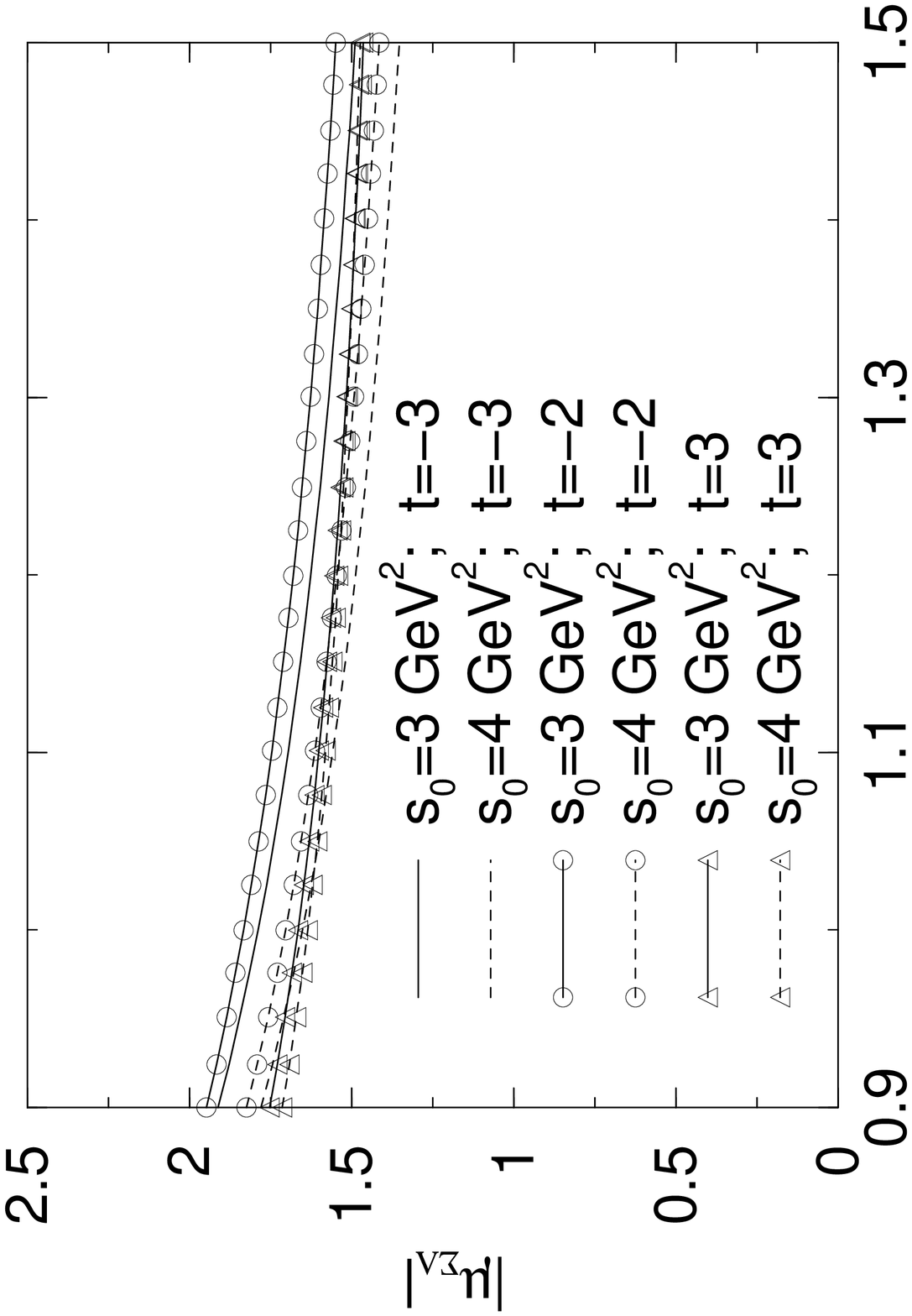}
\vskip 9.0cm
\begin{center}
Figure {\bf 1}
\end{center}
\end{figure}


\begin{figure}  
\vskip 1.5 cm
    \includegraphics{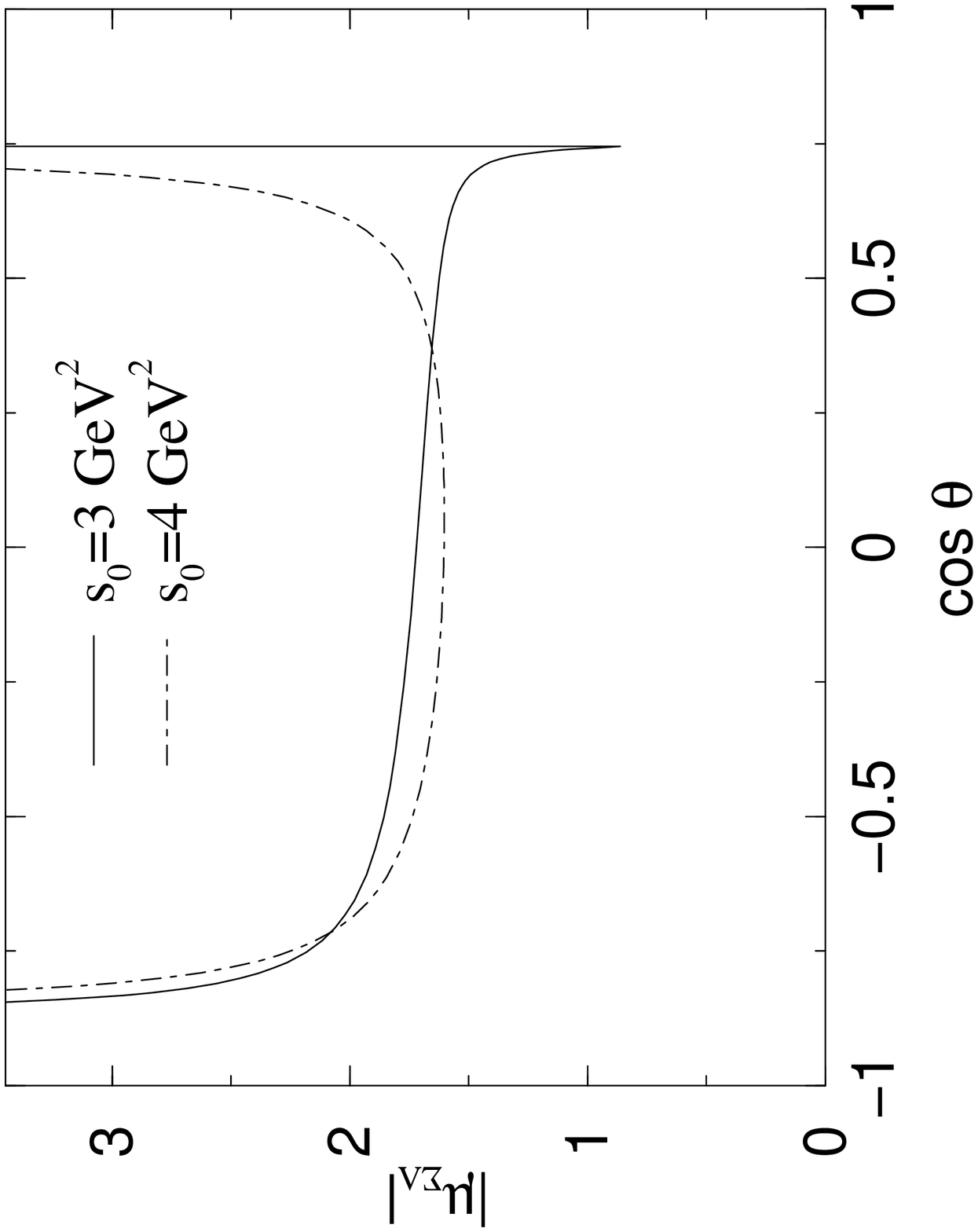}
\vskip 9.5 cm
\begin{center}
Figure {\bf 2}
\end{center}
\end{figure}

\end{document}